\documentclass[sigconf]{acmart}
\usepackage{multicol}
\usepackage{multirow}
\renewcommand{\arraystretch}{1.1}

\AtBeginDocument{%
  \providecommand\BibTeX{{%
    \normalfont B\kern-0.5em{\scshape i\kern-0.25em b}\kern-0.8em\TeX}}}
    
\setcopyright{acmcopyright}
\copyrightyear{2018}
\acmYear{2018}
\acmDOI{XXXXXXX.XXXXXXX}

\acmConference[Conference acronym 'XX]{Make sure to enter the correct
  conference title from your rights confirmation emai}{June 03--05,
  2018}{Woodstock, NY}

\acmPrice{15.00}
\acmISBN{978-1-4503-XXXX-X/18/06}

\begin{document}

\title{MM-GEF: Multi-modal representation meet collaborative filtering}

\author{Hao Wu}
\affiliation{%
  \institution{Huawei Ireland Research Center}
  \city{Dublin}
  \country{Ireland}}
\email{hao.wu7@huawei-partners.com}

\author{Alejandro Ariza-Casabona}
\affiliation{%
  \institution{University of Barcelona}
  \city{Barcelona}
  \country{Spain}}
\email{alejandro.ariza14@ub.edu}

\author{Bartłomiej Twardowski}
\affiliation{%
  \institution{Universitat Autonoma de Barcelona}
  \city{Barcelona}
  \country{Spain}}
\email{bartlomiej.twardowski@huawei.com}

\author{Tri Kurniawan Wijaya}
\affiliation{%
  \institution{Huawei Ireland Research Center}
  \city{Dublin}
  \country{Ireland}}
\email{tri.kurniawan.wijaya@huawei.com}

\begin{abstract}
In modern e-commerce, item content features in various modalities offer accurate yet comprehensive information to recommender systems. The majority of previous work either focuses on learning effective item representation during modeling user-item interactions, or exploring item-item relationships by analysing multi-modal features. These methods, however, fail to incorporate the collaborative item-user-item relationships into the multi-modal feature-based item structure. In this work, we propose a graph-based item structure enhancement method called  MM-GEF: Multi-Modal recommendation with Graph Early-Fusion, which effectively combines the latent item structure underlying multi-modal contents with the collaborative signals. Instead of processing the content features in different modalities separately, we show that the early-fusion of multi-modal features provides significant improvement. MM-GEF learns refined item representations by injecting structural information obtained from both multi-modal and collaborative signals. Through extensive experiments on four publicly available datasets, we demonstrate systematic  improvements of our method over state-of-the-art multi-modal recommendation methods.
\end{abstract}


\maketitle


\section{Introduction}
\label{sec:intro}

The overwhelming prevalence of multimedia content on the Internet has deeply transformed consumer habits of modern e-commerce platforms like Amazon, Taobao, and eBay. Users of these platforms are now increasingly reliant on the multimedia feed. To cope with such changes, recommender systems have also evolved, 
aiming to help users identify items of interest by exploiting multi-modal contents of items (e.g. text descriptions, item images). 
Typically, the raw features of these contents are high-dimensional and therefore difficult to directly incorporate into recommender systems. The prevalence of large-scale pre-trained language and visual models has greatly benefited multi-modal recommender systems by providing low-dimensional modality-specific representations. Some early progress~\cite{vbpr,wang2017your} leveraged pre-trained resources to enrich the item representations. Within this research direction, one particular challenge rises from the huge semantic gap among different modalities, which makes the fusion process a crucial part of the multi-modal recommendation.

The need for jointly exploiting multiple modalities for several downstream tasks has attracted a lot of research attention, specially in the visiolinguistic field. 
Pre-trained visiolinguistic models~\cite{clip, vilbert,lin2023vila} unified embedding spaces for two modalities by jointly training separate text and image encoders. Given their zero-shot performance, pre-trained visiolinguistic models could be used in a multi-modal recommender system to perform early-fusion of text/image representations that belong to a common multi-modal space. 


Inspired by the recent success of graph convolutional neural networks(GCN)~\cite{hamilton2017inductive,gat,welling2016semi},~\cite{LATTICE21} proposes LATTICE for multi-modal recommendation. Instead of directly using multi-modal features as item representation, they focus on mining the latent structure of items for each modality. However, this method fails to incorporate the item relationship derived from collaborative signals, therefore unable to explore the latent structural information underlying the high-order item-user-item 
interactions. Additionally, the item representation in different modalities is only implicitly combined by adding up the edge weights of the homogeneous item graph. We argue that this method can be improved by considering the cross-modal interaction and enabling early multi-modal feature fusion with pre-trained visiolinguistic models. 
Therefore, we propose a graph-based item structure enhancement method for multi-modal recommendation task
to learn the item graph structure by jointly considering all multi-modal and collaborative information.

The main contributions of this paper are as follows:
\begin{itemize}
\item 
 We show the power of early-fusion strategies and pre-trained contrastive visual language representations to guide the multi-modal recommendation modeling task;
\item 
 We propose a novel architecture (MM-GEF) that exploits cross-modal relations and implicit collaborative signals within an early-fused item graph structure and leverages graph neural network modeling capabilities for accurate recommendations;
    \item 
We conduct extensive experiments on four public multi-modal datasets to show the effectiveness of MM-GEF against current state-of-the-art solutions.
\end{itemize}

\section{Related Work}
\label{sec:relatedwork}

In this section, we focus on the previous work related to our proposed method, 
including the topics of multi-modal learning and multi-modal recommendation.

\subsection{Multi-modal Learning}
User decisions are dependent on how they process distinct data sources coming from a multi-modal environment. 
To match this multi-modal behaviour, multiple approaches have been proposed to fuse multi-modal signals and leverage fused representations in downstream tasks.
Thanks to the modeling capabilities of Transformer-like models, there is an emerging trend of derivative models for multi-modal learning \cite{uniter,vilbert,clip,vlbert} that commonly perform intermediate fusion at different network stages for gradual joint representations. For instance, CLIP~\cite{clip} provides pre-trained image and text encodings belonging to the same multi-modal latent space that can be used for early fusion in our recommender system while still mitigating any semantic gap between different modalities.

\subsection{Multi-modal Recommendation}
Existing studies are mainly based on the early success of collaborative filtering (CF) methods \cite{cf-usenetnews-grouplens,cf-netnews-grouplens} adapted to account for multi-modal data as side information, thus enriching the item representations~\cite{jia2015multi,min2015cross,singh2008relational}. The benefits of such models are that they are able to leverage content-based item representations from early stages of model training and do not simply rely on implicit user-item interaction signals.
Recently, MM-GCN \cite{MMGCN} modeled modal-specific user preferences by applying an intermediate-late modality fusion. Furthermore, GRCN\cite{grcn} used multi-modal information to remove noisy edges from the collaborative graph. MML~\cite{pan2022multimodal} incorporates multi-modal side information in a sequential recommender system, by combining multi-modal pre-trained features with item ID  embedding using a self-attention block. Similarly, ~\cite{liu2021pre} injects multi-modal content features into a homogeneous item graph and pretrain the model by reconstructing both graph structures and node content features. Given the power of pre-trained visiolinguistic models for multi-modal understanding, MM-Rec \cite{mmrec} uses ViLBERT~\cite{lu2019vilbert} encoder to feed modality-aware representations to their crossmodal recommendation architecture.
 MVGAE~\cite{yi2021multi} proposed a multi-modal variational graph auto-encoder model that learns Gaussian variables for nodes, representing semantic information and uncertainty. 
These proposed methods showed that multi-modal signals can contribute to the recommender system by incorporating content features as side information. However, they have two major limitations: first, even though multi-modal features complement item representations, they heavily rely on collaborative signals, thus suffering from data sparsity. 
Second, in graph-based models, noisy item content features may add up to noisy implicit signals and lead to erroneous graph signal propagation and suboptimal performance.

Instead of linking multi-modal content features to item representation, another direction consists of exploring item structural information using the rich semantic information contained in content features. 
KGAT \cite{KGAT19} utilizes a knowledge graph and builds connections between items based on their attributes. 
Considering that items are associated with rich contents in multiple modalities, 
LATTICE~\cite{LATTICE21} argues that the latent semantic item-item structures underlying these multi-modal contents could be beneficial for learning better item representations and 
proposed a gradual fusion architecture. 

However, LATTICE has inherent limitations as it does not leverage high order collaborative signals during the construction of the item graph, missing out on the rich contextual information provided by user-item interactions. Additionally, it leverages multi-modal content features separately, neglecting the critical cross-modal interactions at an early stage.

\section{Methodology}
\label{sec:method}

\begin{figure*}[t]
    \centering
    \includegraphics[width=1\textwidth, keepaspectratio]{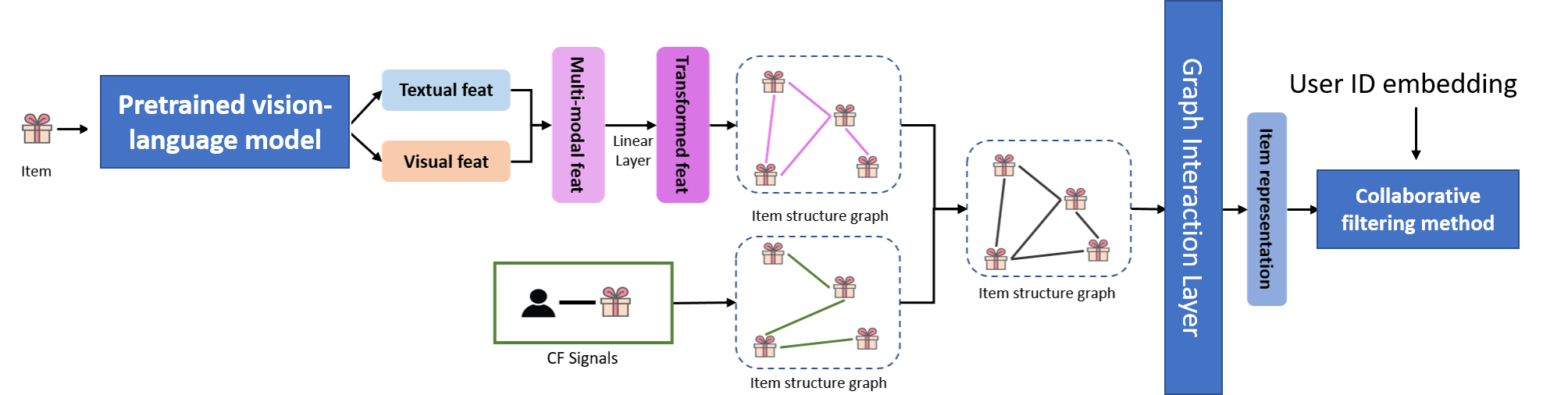}
    \caption{Architecture illustration of MM-GEF}
    \label{fig:model_architecture}
\end{figure*}

We illustrate the general architecture of MM-GEF in Figure~\ref{fig:model_architecture}.
It first learns item structural graphs from both multi-modal features and collaborative signals and perform graph convolution propagation on the item graph to explore high-order item-item relationships. 
The refined item representations obtained from the graph interaction network are then directly used with item embeddings in the downstream
CF method.


\subsection{Multi-modal feature extraction}
One contribution of our work is to recognize that the item content features in various modalities are related. Instead of processing multi-modal features independently~\cite{LATTICE21,MMGCN} and only fusing different modalities before decision-making, early-fusion methods can capture the cross-modal information more effectively. Our proposed early-fusion strategy is implemented by extracting content features from a pre-trained vision-language model, for example CLIP~\cite{clip}. Recall that CLIP is trained on the cosine similarity of text and image pairs. Based on this setting we conduct early multi-modal fusion by averaging the pre-trained visual and textual features. 

\subsection{Item structure learning} 
Considering the fact that the raw content feature could be noisy and of high dimensionality, we propose to 
compress
the item multi-modal feature in order to have effective and efficient item representations.
During training of the model, the CLIP feature will be transformed into the low-dimensional latent space with a trainable linear layer.

Following~\cite{LATTICE21}, we build a homogeneous item graph by considering the cosine similarity of item pairs. In detail, for each item $i$, we build weighted edges for the top-k most similar items based on cosine similarity score of the content feature: 
\begin{equation}
    M_{ij} = CosineSim(I_i,I_j)
\end{equation}
\begin{equation}
     E_{ij}^m =
    \begin{cases}
      M_{ij}  & \text{if $M_{ij}$ $\in$ topk($M_i$)}\\
      0 & \text{otherwise}
    \end{cases}
\end{equation}
$M_{ij}$ is the affinity score of the item pair $i$ and $j$, and $E^m$ is the graph edge set.

\subsection{Item structure aggregation layer}
In addition to learning from multi-modal features, item structural information can also be gained from user-item interaction. We propose to incorporate the collaborative signals in augmenting the quality of the item-item structure. We build another homogeneous item graph based on item-user-item relations as: 
\begin{equation}
    C_{ij} = \sum_{u \in U} u_i \& u_j
\end{equation}
\begin{equation}
    E_{ij}^c = \frac{C_{ij}}{C_{max}}
\end{equation}
where $u_i$ indicates the interactions between user $u$ and item $i$, $C_{ij}$ is the number of times that both item $i$ and $j$ interacted with a same user. Next, we use normalized $C_{ij}$ as the edge weight.
The item-item graphs based on multi-modal feature($E^m$) and collaborative signals($E^c$) are then combined as a weighted sum of both $E^m$ and $E^c$ using soft attention mechanism: 
\begin{equation}
    E^u = SoftAttn(E^m:E^c)
\end{equation}
where $E^u$ is the edge set of the unified item structure graph.

\subsection{Graph Interaction Layer}
We use a graph interaction network to inject high-order item-item relationships:
\begin{equation}
    h_i^t = \sum_{j \in N_i}^{}E^u_{ij}h_i^{t-1}
\end{equation}
where $h_i^t$ is the hidden state of the item $i$ in step $t$. We use item id embedding to initialize the graph interaction network.

\subsection{Joining with CF methods} After obtaining the item representations learned from item structure, we follow ~\cite{LATTICE21} and combine them with downstream CF method. In detail, we normalize the learned item representation $h_i$ and directly add it to the item embedding of a CF method. Lastly, the user-item recommendation score is computed as the inner product of user ID embedding and refined item embedding.

\section{Experiments}

\subsection{Datasets, experimental setting, metrics}
We conducted experiments with four publicly available datasets: three categories of the widely used Amazon dataset~\cite{mcauley2015image} and the MovieLens 10M dataset~\cite{harper2015movielens}. Datasets from Amazon's categories were selected and created following the procedure from other multi-modal recommender works~\cite{LATTICE21}. Additionally, the MovieLens 10M was created by taking trailer video key frames and text descriptions of the movies (title, description) from the IMDB website. Then, we use 768-dimensional visual and textual features from the pre-trained CLIP model~\cite{clip}. 


For performance comparison we use: Recall@N, Precision@N, and NDCG@N metrics that assess ranking quality. Following other works~\cite{he2020lightgcn,MMGCN,LATTICE21}, we use $N=20$ as a default value and report the average for all metrics after 5 different runs on the test split. All results are presented as percentages.

\subsection{Implementation details}
We use CLIP~\cite{clip} to extract pretrained visual and text features, GCN as graph interaction network and LightGCN~\cite{he2020lightgcn} as downstream CF method.
We search batch size in $\left [ 512, 1024, 2048, 4096 \right ]$, and learning rate in $\left [ 1e-2, 1e-3,1e-4 \right ]$. The transformed multi-modal feature embedding size, the size of the hidden state of GCN, item embedding and user embedding are all set to 64, the number of neighbors ($k$) during structure learning is 10.
For the cold start setting, lr=$1e-5$ is used. ~\footnote{Code will be publicly available upon acceptance.}

\begin{table*}[h]
\caption{Methods comparison and ablation study of MM-GEF on different datasets. Scores are presented as percentages.}
\label{tab:comparison}
\setlength{\tabcolsep}{3pt}
\centering
\resizebox{0.93\textwidth}{!}{%
\begin{tabular}{l|ccc|ccc|ccc|ccc} 
\hline
\multirow{2}{*}{\textbf{Model}} & \multicolumn{3}{c|}{\textbf{Amazon Baby}}           & \multicolumn{3}{c|}{\textbf{Amazon Sports}}        & \multicolumn{3}{c|}{\textbf{Amazon Clothing}}                   & \multicolumn{3}{c}{\textbf{MovieLens-10M}}                      \\
 & R@20 & P@20 & N@20 & R@20 & P@20 & N@20 & R@20 & P@20 & N@20 & R@20 & P@20 & N@20 \\ 
\hline
CB                  & 1.44 & 0.07 & 0.56 & 2.08 & 0.11 & 0.86 & 3.38 & 0.17 & 1.40 & 1.19 & 0.30 & 1.97 \\
LightGCN            & 7.21 & 0.38 & 3.19 & 7.83 & 0.41 & 3.66 & 6.73 & 0.17 & 2.61 & 33.32 & 7.44 & 26.23 \\
LATTICE             & 8.19 & 0.43 & 3.66 & 8.90 & 0.47 & 4.11 & 6.90 & 0.35 & 3.13 & 	\textbf{33.98} & 7.64 & 36.99 \\
\hline
MM-GEF w/o CF+Att   & 8.12 & 0.43 & 3.58 & 8.89 & 0.47 & 4.12 & 6.93 & 0.35 & 3.21 & 33.46 & 7.45 & 24.01 \\
MM-GEF w/o CF       & 8.08 & 0.43 & 3.61 & 8.66 & 0.46 & 4.02 & \textbf{8.78} & 0.46 & 4.05 & 33.05 & 7.38 & 23.61 \\
MM-GEF w/o Att      & 8.21 & 0.43 & 3.65 & 9.41 & 0.49 & 4.25 & 7.71 & 0.39 & 3.50 & 33.30 & 7.44 & 23.99 \\
	\textbf{MM-GEF}     & 	\textbf{8.69} & 	\textbf{0.46} & 	\textbf{3.83} & 	\textbf{9.54} & 	\textbf{0.50} & 	\textbf{4.26} & 	8.31 & 	\textbf{0.42} & 	\textbf{3.76} & 33.98 & 	\textbf{7.64} & 	\textbf{37.11} \\
\hline
\end{tabular}
} 
\end{table*}
\subsection{Methods comparison and ablation study}
We compare MM-GEF method against three baseline methods: 
\begin{itemize}
    \item LightGCN~\cite{he2020lightgcn} - simple and efficient approach with GCN and layer combination.
    \item LATTICE~\cite{LATTICE21} - our main baseline where the downstream CF task is LightGCN.
    \item CB - content-based recommendation based on text and image similarity as side information.
\end{itemize} The results are presented in Table~\ref{tab:comparison}. As expected, LATTICE is systematically better than LightGCN itself. For all Amazon datasets, MM-GEF shows better performance than the baseline methods in all metrics. Interestingly, for the biggest dataset used in this experiment MovieLens-10M, Recall@20 is better for LATTICE. However, in the meaning of ranking precision (NDCG and Precision), MM-GEF gives the best outcome. Recommendations based on content alone are far behind.
\vspace{-1em}
\paragraph{\textbf{Ablation study:}} In Table~\ref{tab:comparison}, we present the MM-GEF method without the novel elements described in Section~\ref{sec:method}: early fusion with CF, soft-attention (Att), and early fusion with CF and soft-attention (CF+Att). If we compare the MM-GEF w/o CF+Att we can see the gain from using CLIP features in comparison to the LATTICE method. The results are comparable for the Amazon dataset, with a big difference for NDCG@20 and the MovieLens-10M dataset. Soft-attention alone (MM-GEF w/o CF) improves for Amazon Clothing and NDCG@20 for Amazon Baby. Early fusion with CF (MM-GEF w/o Att) improves the results significantly for all datasets, except MovieLens-10. The non-ablated version of MM-GEF produces the best results.

\begin{table*}
\caption{Modality comparison on multiple datasets.}
\label{tab:mm}
    \centering
    \setlength{\tabcolsep}{3pt}
\resizebox{0.93\textwidth}{!}{%
\begin{tabular}{l|ccc|ccc|ccc|ccc}
\hline
& & & & \rule{1.2cm}{0pt} & \rule{1.2cm}{0pt} & \rule{1.2cm}{0pt} & \rule{1.2cm}{0pt} & \rule{1.2cm}{0pt} & \rule{1.2cm}{0pt} & \rule{1.3cm}{0pt} & \rule{1.2cm}{0pt} & \rule{1.2cm}{0pt} \\[-\arraystretch\normalbaselineskip]
     \multirow{2}{*}{\textbf{Model}} &    \multirow{2}{*}{\textbf{CF}} &  \multirow{2}{*}{\textbf{Text}} &   \multirow{2}{*}{\textbf{Img}} & \multicolumn{3}{c|}{\textbf{Amazon Baby}} & \multicolumn{3}{c|}{\textbf{Amazon Sports}} & \multicolumn{3}{c}{\textbf{Amazon Clothing}} \\
    & & & & R@20 & P@20 & NDCG@20 & R@20 & P@20 & NDCG@20 & R@20 & P@20 & NDCG@20 \\
\hline
   \multirow{2}{*}{LATTICE} & & \checkmark & \checkmark & 8.13 & 0.43 & 3.58 & 8.93 & 0.47 & 4.10 & 6.88 & 0.35 & 3.19 \\  & \checkmark & \checkmark & \checkmark & 8.2 & 0.44 & 3.66 & 9.41 & 0.50 & 4.25 & 7.71 & 0.39 & 3.50 \\
\hline
\multirow{6}{*}{MM-GEF} & & & \checkmark & 8.28 & 0.44 & 3.63 & 8.73 & 0.46 & 4.01 & 4.29 & 0.22 & 1.98 \\
 & & \checkmark & & \textbf{8.71} & 0.46 & 3.80 & 8.95 & 0.47 & 4.06 & 7.32 & 0.37 & 3.33 \\
 & & \checkmark & \checkmark & 8.57 & 0.45 & 3.76 & 8.87 & 0.47 & 4.11 & 7.55 & 0.38 & 3.41 \\
 & \checkmark & & \checkmark & 8.71 & \textbf{0.46} & \textbf{3.82} & \textbf{9.56} & \textbf{0.50} & 4.26 & 7.99 & 0.41 & 3.60 \\
 & \checkmark & \checkmark & & 8.67 & 0.46 & 3.81 & 9.32 & 0.49 & 4.20 & 8.29 & 0.42 & 3.76 \\
 & \checkmark & \checkmark & \checkmark & 8.64 & 0.46 & 3.77 & 9.54 & 0.50 & \textbf{4.26} & \textbf{8.31} & \textbf{0.42} & \textbf{3.76} \\
\hline
\end{tabular}
} 
\end{table*}

\subsection{Effect of different modalities}
Table~\ref{tab:mm} presents model performances with various combinations of modalities. We run LATTICE with and without CF, and all allowed combinations for MM-GEF. It can be seen that each dataset has its own characteristics. For Amazon Baby and Sports, collaborative filtering with visual features from images results in good performance. On the Amazon Clothing dataset, MM-GEF benefits from all modalities. LATTICE in all datasets and metrics performs far behind MM-GEF.

\subsection{Cold-start setting experiments}
In order to simulate the cold-start problem, we randomly selected 20\% of items and removed all user-item interaction pairs associated with these items from the training set. The test set contains all items, so those removed 20\% are unseen during training. The results for the Amazon Baby dataset and different modalities are presented in Table~\ref{tab:cold}. The best overall method is MM-GEF with all modalities. In comparison to LATTICE\footnote{The results for LATTICE are slightly different from the original work~\cite{LATTICE21}, as 20\% random items are sampled as instructed, but we couldn't replicate their exact selection.} the results are significantly better.

\begin{table}
\caption{Cold-start setting results for Amazon Baby dataset. Best results in bold and second-best results with underlines.}
\label{tab:cold}
\centering
\resizebox{\columnwidth}{!}{%

\begin{tabular}{l|ccc|ccc}
\bottomrule
 & \rule{.7cm}{0pt} & \rule{.7cm}{0pt} & \rule{.7cm}{0pt} & \rule{1.1cm}{0pt} & \rule{1.1cm}{0pt} & \rule{1.1cm}{0pt} \\[-\arraystretch\normalbaselineskip]
 Model & CF & Text & Img & R@20 & P@20 & NDCG@20 \\
\hline
   LATTICE & \checkmark & \checkmark & \checkmark & 1.00 & 0.07 & 0.51 \\
\hline
\multirow{6}{*}{MM-GEF} & & & \checkmark & 0.67 & 0.05 & 0.39 \\
 & & \checkmark & & 0.84 & 0.06 & 0.46 \\
 & & \checkmark & \checkmark & 0.84 & 0.06 & 0.46 \\
 & \checkmark & & \checkmark & \underline{1.48} & \underline{0.11} & \underline{0.71} \\
 & \checkmark & \checkmark & & 1.41 & 0.10 & 0.70 \\
 & \checkmark & \checkmark & \checkmark & \textbf{1.51} & \textbf{0.11} & \textbf{0.74} \\
\toprule
\end{tabular}
}
\end{table}

\section{Conclusion and Future Work}
\label{sec:conclusions}
In this work we propose a novel method MM-GEF for multi-modal recommendations where graph-based item structure based on different modalities is enhanced by early fusion with signals from collaborative filtering. In a series of experiments on four real-life datasets, MM-GEF effectively exploits multi-modal recommendation and achieves better results than current state-of-the-art methods in almost all cases. The ablation study shows that all components of MM-GEF are necessary to achieve the best performance. Moreover, MM-GEF demonstrates its superiority in the cold-start situation. Extending the MM-GEF method still can be extended by using more modalities (e.g. audio, video), better pre-trained encoders, or graph structure modeling that helps with an early-fusion approach.

\clearpage

\newpage
\bibliographystyle{ACM-Reference-Format}
\bibliography{main}

\end{document}